# Status and Perspectives of Neutrino Physics

A. Bettini. Dipartimento di Fisica G. Galilei dell'Università di Padova and INFN Sezione di Padova

I will first give a brief but comprehensive review of the status of our knowledge in neutrino physics. With reference to a not too far future I will then discuss the perspectives that appear to me to be most important and promising.

## 1. Introduction

We know three kinds of neutrinos, $\nu_e$, $\nu_\mu$ and $\nu_\tau$, with the flavour lepton numbers of the electron, muon and tauon respectively. These flavour neutrino states are usually classified to belong to the first, second and third family, but, contrarily to the other members, neutrinos have the unique property to change their flavour - hence family - in time.

Neutrinos differ from the other elementary constituents of matter from another point of view. The antiparticles of each quark and charged lepton are distinguished from the particles because all their charges have opposite values. On the contrary, neutrinos have no electric charges and the lepton number might not be absolutely conserved, implying that neutrinos and antineutrinos may be two states of the same particle (Majorana neutrinos).

Neither the first neither the second property is compatible with the standard model; the first has already been experimentally observed, as we will discuss immediately, the second not yet. The way is the search for the neutrinoless double beta decay ($0\nu 2\beta$), discussed in §4.

## 2. Neutrino mass spectrum and mixing

Neutrinos have been observed to change flavour in two different ways.

The first phenomenon is the "neutrino oscillations", which, in its purest form, takes place in a vacuum. As such, it is purely kinematical, formally described by the kinetic part of the Hamiltonian. It has been discovered as the disappearance of the muon neutrinos indirectly produced by cosmic rays in the atmosphere. Typical neutrino energies ($E$) are from sub-GeV to multi-GeV, while the flight lengths ($L$) are up to a few thousands kilometres. The phenomenon has been confirmed by the K2K experiment with an accelerator $\nu_\mu$ source ($E \approx 1$ GeV) on a $\approx 200$ km flight length.

The probability to observe the flavour state $\nu_\beta$ in a beam initially pure $\nu_\alpha$ (monoenergetic, with energy $E$) contains oscillating terms of the type

$$P_{\alpha\beta} = A(\theta_{12}, \theta_{23}, \theta_{13}) \sin^2(1.27(m_i^2 - m_j^2)L/E)$$

where $m$ is in eV, $L$ in km and $E$ in GeV.

The oscillation amplitude $A$ is a function of the three mixing angles, $\theta_{ij} \in [0, \pi/2]$. The oscillation frequencies are proportional to the differences between the squares of the masses of the eigenstates.

One sees that the phenomenon is independent on the sign of $m_i^2 - m_j^2$. Furthermore, the expressions of the amplitudes, not shown here, are symmetrical under the reflection through 45°, namely independent on $\text{sgn}(\pi/2 - \theta_{ij})$.

The second phenomenon is the flavour conversion in matter, the Mikehev-Smirnov-Wolfenstein[1] effect. It is a dynamical phenomenon, due to the $\nu_e e$ interaction potential in the matter. Indeed, the neutral current (NC) interaction is equal for all the neutrino flavours and does not distinguish amongst them. Electron neutrinos interact via charged currents (CC) both with the nucleus and the electrons. But it easy to see that the weak potentials due to the protons and the neutrons cancel and only the interaction with electrons gives a net result.

The potential depends on neutrino energy and matter density, and has opposite sign for neutrinos and antineutrinos. While crossing a variable density medium $\nu_e$'s may reach a critical density layer (level crossing), where a resonant transition to a coherent superposition of the other flavours may happen. The phenomenon is the dominant process in the Sun, for neutrino energies above about 1 MeV (and is expected to happen in the Earth and in the Supernovae). It corresponds to a square mass difference, call it $\delta m^2$, much smaller than the "atmospheric" one. The matter effect does depend both on the $\text{sgn}(\delta m^2)$ and $\text{sgn}(\pi/2-\theta_{12})$. The flavour conversion at the scale $\delta m^2$ has also been observed as an oscillation (in vacuum) in the KamLAND experiment on $\bar{\nu}_e$ from reactors (see later).

This evidence tells us that the flavour states are not stationary, they are not the (mass) eigenstates. We call these last $\nu_1$, $\nu_2$, and $\nu_3$, and $m_1$, $m_2$ and $m_3$ their masses. Flavour states are linear combinations of the eigenstates, $\nu_l = \sum_{i=1}^{3} U_{l,i}\nu_i$, where $l = e, \mu, \tau$ and $U$ is the mixing matrix. $U$ is unitary if, as we will assume, the eigenstates are orthogonal. We can then write $U$ as the product of three rotation matrices including a phase factor, as in the case of quarks, and a fourth diagonal matrix with two more phases (Majorana phases). The last can be absorbed in the wave functions only if neutrinos and antineutrinos are different particles (Dirac neutrinos). In general we have

$$U = \begin{pmatrix} 1 & 0 & 0 \\ 0 & c_{23} & s_{23} \\ 0 & -s_{23} & c_{23} \end{pmatrix} \begin{pmatrix} c_{13} & 0 & s_{13}e^{i\delta} \\ 0 & 1 & 0 \\ -s_{13}e^{-i\delta} & 0 & c_{13} \end{pmatrix} \times$$

$$\times \begin{pmatrix} c_{12} & -s_{12} & 0 \\ s_{12} & c_{12} & 0 \\ 0 & 0 & 1 \end{pmatrix} \begin{pmatrix} 1 & 0 & 0 \\ 0 & e^{i\alpha} & 0 \\ 0 & 0 & e^{i\beta} \end{pmatrix}$$

where $c_{ij} = \cos\theta_{ij}$, and $s_{ij} = \sin\theta_{ij}$

In total there are nine quantities to measure, three masses, three mixing angles and the three phases. The last, if $\neq 0$ and $\neq \pi$, give CP violation effects in the lepton sector. Majorana phases are, even if present, irrelevant in oscillation and matter conversion phenomena; they appear only in double beta decay (and similar phenomena, too small to be detected).

Different groups[2] have performed global fits including all the available data, providing values for the three mixing angles (only an upper limit for $\theta_{13}$) and two squared mass differences. Results can be summarised as follows[2a]:

$\delta m^2 \equiv m_2^2 - m_1^2 = 83 \pm 3$ meV$^2$
$|\Delta m^2| \equiv |m_3^2 - m_2^2| \approx |m_3^2 - m_1^2| = 2400 \pm 300$ meV$^2$
$\theta_{12} = 33°\pm 2°$; $\theta_{23} = 45°\pm 3°$; $\theta_{12} < 10°$ (2$\sigma$)

We can now define more precisely the eigenvalues. We count them in order of decreasing content of $\nu_e$: $\nu_1 \approx 70\%$ $\nu_e$, $\nu_2 \approx 30\%$ $\nu_e$, $\nu_3 <$ few% $\nu_e$. Fig. 1 shows the spectrum. We do not know whether the singlet is above (so called normal hierarchy) or below (inverse hierarchy) the doublet. Neither we know the absolute scale; we have only the lower limits $m_3 > \sqrt{\Delta m^2} \approx 50$ meV in the first case, $m_1, m_2 > \sqrt{\Delta m^2} \approx 50$ meV in the second.

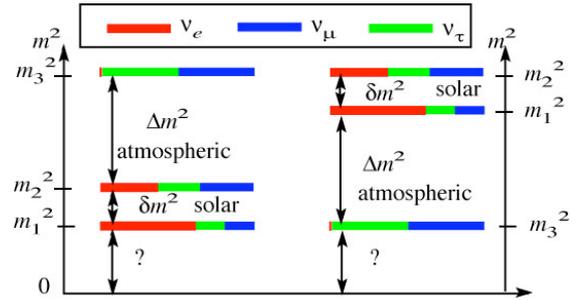

Fig.1 Neutrino mass spectrum

Notice that the ratio of the two square mass differences $\alpha = \delta m^2/\Delta m^2 = 0.03$ is, as anticipated, small; being $|\theta_{13}|$ small too, the two phenomena are almost decoupled.

Solar neutrino experiments have provided the following data: the fluxes above the energy thresholds (different in the different experiments), the energy spectra during the day and during the night, the ratio NC/CC rates from SNO. Historically, these data defined a set of solutions (Vacuum, quasi-vacuum, LOW, SMA, LMA) that can represented as regions in the parameter space $\delta m^2$ versus $\sin^2\theta_{12}$. After the release of the SNO data only the so-called Large Mixing Angle (LMA) solution remains. Additional data are from the KamLAND experiment, a liquid scintillator detector located in the Kamioka mine in Japan. It measures the $\bar{\nu}_e$ flux and, even more important, the energy spectrum from the power reactors, with a dominant baseline of 180 km. The Collaboration has delivered the latest results for an exposure of 766 t yr (previous was 162 t yr) at the Neutrino 2004 Conference in June[3]. The results are consistent with the LMA solution and strongly improve the resolution on $|\delta m^2|$. Notice that solar neutrinos and reactor antineutrinos give the same solution, a fact that provides the first test of CPT invariance in the neutrino sector. With further data, KamLAND will still improve a bit the resolution on $|\delta m^2|$.

The knowledge of $\theta_{12}$ is dominated by solar data. We can still expect a slight improvement in the future mainly from CC/NC data from SNO, provided this quantity is measured at a few percent level.

The fit of J. Bahcall et al.[2b] includes as free parameters the ratios between fitted fluxes and those of the solar standard model (SSM)[4] $f_i = \Phi_i/\Phi_{SSM}$, with the only constraint of the solar luminosity, obtaining $f_{pp}=1.01\pm0.02$, $f_{7Be} = 1.03^{+024}_{-1.03}$, $f_{8Bo}=0.87\pm0.0$, , and $f_{CNO} = 0.0^{+2.7}_{-0.0}$. We see that the $pp$ and $^8$B fluxes are well determined, independently on the solar model; while $^7$Be and the (small) CNO are poorly known. BOREXINO[5] at LNGS will soon accurately measure the $^7$Be flux.

Before leaving this subject, let me notice that if the mixing angle is just $\theta_{12}=45°$, the MSW transition is not effective. That value being now excluded by the solar data at 5.8 $\sigma$, the existence of the MSW effect is proven.

Our knowledge of $|\Delta m^2|$ and $\theta_{23}$ is due to atmospheric neutrinos, specifically the disappearance of $\nu_\mu$'s on distances comparable with the Earth radius. The data set is dominated by the SuperKamiokande experiment. There have been no substantial new data in the last year, but new analyses with a better modelling of the source. Improvements include three-dimensional modelling of the fluxes and improved cross-section values to better agree with K2K[7] near detector. Consider for example the results of Gonzalez-Garcia and Maltoni[2c]. Comparing the old and the new fits for the three SuperK samples of Sub-GeV, multi-GeV and up-going $\mu$, the best values of $|\Delta m^2|$ decrease by 200 meV$^2$, 800 meV$^2$ and 500 meV$^2$ respectively. The example shows that the results are dominated by systematic uncertainties. As the authors observe, the main ones are the flux energy independent normalisation and the uncertainty in its energy dependence, heuristically parameterised as $E^{-\gamma}$. The uncertainties on the cross sections are sizeable, but less important. $|\Delta m^2|$ and its uncertainty can be pinned down by the long base line experiments with man-made sources at accelerators. K2K is giving the first contributions, but its luminosity is small. Improvement is expected from the soon-to-come NUMI+MINOS[7] at Fermi Lab and CNGS, OPERA & ICARUS[8] at CERN+LNGS.

For the last mixing angle $\theta_{13}$, CHOOZ[9], a disappearance experiment on $\bar{\nu}_e$ from (two) reactors (1 km baseline), provides the limits $|\theta_{13}|<10°$ or $|\theta_{13}|>80°$; solar data chose the first solution.

## 2. Eight challenges in neutrino physics

Speculating in the not too far future, in the next dozen of years or so, I believe that the principal challenges will be

1. Establish the nature of the neutrinos, Majorana or Dirac, through the search for $0\nu2\beta$ decays. Searching in several isotopes is mandatory, given the large uncertainty in the nuclear matrix elements.
2. Complete the ongoing program (and look for surprises)
   a. Is the majority "atmospheric" $\nu_\mu$ oscillation really dominantly into $\nu_\tau$? CNGS ($\nu_\tau$ appearance), NUMI+MINOS indirectly
   b. Complete the measurement of the solar neutrino spectrum components. $^7$Be by BOREXINO and KamLAND (solar phase); accurate measurement of $pp$ flux should be at a few percent accuracy (LENS and other proposals)
3. How small is $\theta_{13}$?
   a. $\nu_\mu \rightarrow \nu_e$ from accelerators (NUMI, and CNGS will give small improvement, T2K[10] substantial)
   b. Disappearance of reactor $\bar{\nu}_e$'s (many proposals world wide; the technique is dominated by systematic; the feasibility has not been shown yet)
4. Improve precision on $\theta_{12}$ and $\theta_{23}$: NUMI off-axis etc. Measure sgn($\theta_{23}-\pi/2$) through matter effects.
5. Absolute scale of mass; beta decay, $0\nu2\beta$ decay, cosmology.
6. Sign of $\Delta m^2$; Supernova neutrinos, Atmospheric neutrinos [needs >100 kt magnetised tracking calorimeter a la MONOLITH (INO proposal in India)], high intensity artificial beams, $0\nu2\beta$ experiments
7. CP violation in the Dirac sector. Needs very high intensity beams; it is impossible if $|\theta_{13}|$ turns out to be too small.

8. CP violation in the Majorana sector. $0\nu\beta\beta$ experiments; needs much technological development, a lot of luck and much better knowledge of nuclear matrix elements

## 3. The Cosmic Connection

Cosmology has made tremendous progress in the last several years both in the modelling and in the quantity and, more important, the quality of the observational data. The basic parameters of the model have been consistently determined with good accuracy. But still, the present "standard model" is purely phenomenological and, in particular, the set of basic parameters is not completely defined.

With this caveat, cosmology provides a potentially very sensitive, albeit indirect, means of measuring or limiting the absolute neutrino mass. The relevant property of neutrinos is that, given the smallness of their mass, they are not confined in the large-scale structures of the Universe. Free streaming above them, they tend to erase the structures at scales smaller than a certain value, by amounts dependent on the neutrino mass. For sub-electronvolt masses the suppression happens below a few tens of Mpc. More specifically, we can extract a limit on (or a value of) the fraction of matter density due to neutrinos $f_\nu=\Omega_\nu/\Omega_m$. Knowing $\Omega_m$ (it is known within 15%, a rather large uncertainty compared to those of other cosmic parameters) we have $\Omega_\nu$, which, in turn, gives the sum of the neutrino masses through the relation $\sum m_i$ (eV) $= 94h^2\Omega_\nu$, where $h^2 \approx 0.5$ is the reduced Hubble constant squared.

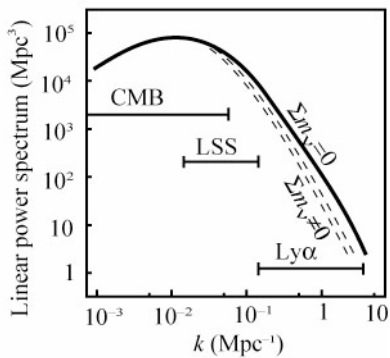

Fig. 2 Sketch of the mass power spectrum

The relevant measured quantity is the large scale structures power spectrum $P(k)$, which is the Fourier transform of the correlation function between two "point" masses, the galaxies at such huge scales (the probability to find two such objects at a distance $d$, over that for a random distribution. $k$ is the variable conjugate of $d$). The function is schematically shown in Fig.2,

$P(k)$ can be determined with different kinds of observations. Presently, three are the most important sources of data: 1. the CMB anisotropies[11] that correspond to early epochs and extremely large scales (Gpc to $\approx 30$ Mpc); at these scales the spectrum is almost insensitive to neutrino mass. Nonetheless, the experiments that measured with high angular resolution the CMB anisotropies, the balloon-born BOOMERANG, MAXIMA and ARCHEOPS, the ground-based DASI and later and more precisely the satellite experiment WMAP have been very important to determine the other cosmological parameters; 2. the large scale structures (LSS) galaxies spectrum (at later epochs) at intermediate scales (100 Mpc to several Mpc; the 2dFGRS[12] and SDSS[13] surveys), rather sensitive to sub-electronvolt neutrino masses; 3. the Lyman alfa forest[14] data at still lower scales (< 10 Mpc) and, for this reason, very sensitive, in principle, to neutrino mass.

Present data do not give any robust evidence for non-zero neutrino masses, but the upper limits are extremely low, the best we have. One must be careful, because the limit depends on the set of assumed basic parameters and on the set of data included in the fit. Degeneracies are present between some parameters, making the results somewhat dependent on the assumed priors. The second issue is mainly the inclusion or not of the Ly-$\alpha$ data. These data are at the scales were the effect of neutrino mass is largest, but the extraction of the correlation function from the data is not completely straightforward.

Typical analysis including CMB and LSS, but not Lyman alfa, give $\sum m_i$ (eV) <2100 meV[15]. From oscillations we know that at the limit the three masses are almost equal. Hence $m_i$<700 meV. More recent analysis, including new results from SDSS and Ly-$\alpha$ forest give $m_i$<130 meV [16] and $m_i$<157 meV [17]. The situation is sketched in Fig.3.

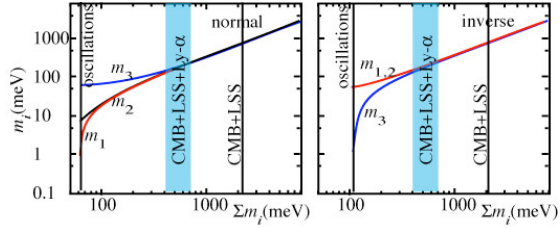

Fig.3. Neutrino masses vs. sum. Dotted lines show schematically the effect of increasing neutrino mass

In consideration of the constant and rapid progress of cosmology both in modelling and in the richness and systematic and statistic accuracy of the data, we can expect further improvements soon. Cosmology might well be close to detecting neutrino mass.

The classic "direct" measurement of the "electron neutrino" mass is based on the search of a distortion very near to the end-point of the electron spectrum from the tritium beta decay $^3H \rightarrow ^3He + e^- + \bar{\nu}_e$. To be precise, mass is a property of an eigenstate, not of $\nu_e$, and what is limited (or measured) is a weighted average of the masses $m_{\nu_e}^2 = |U_{e1}|^2 m_1^2 + |U_{e2}|^2 m_2^2 + |U_{e3}|^2 m_3^2$. Present limit is $m_{\nu e}<2.2$ eV from the Mainz[18] and the Troitz[19] experiments.

In the future the two groups, joining forces, will build a new big spectrometer, KATRIN[20], aiming to reach $m_{\nu e} \approx 200$ meV eV. Even if this value is at the sensitivity of cosmology already now, it will come directly from a laboratory experiment and, as such, will be extremely important. Notice also that it will be sensitive to the signal level claimed by Klapdor et al. in $0\nu 2\beta$ [21].

## 4. The relationship between neutrinos and antineutrinos

Quarks and charged leptons are Dirac particles, for neutrinos we do not know. The charge conjugate of the neutrino might be the neutrino itself (Majorana neutrino). We might call it *Majotrino*. In this case $0\nu 2\beta$ decay may happen, violating the lepton number by two units. The process is so rare that experiments not only require deep underground laboratories but extreme care in reducing the backgrounds due to radiocontaminats. Background is everywhere, in particular in the detector materials and in its surroundings. The struggle for lower mass sensitivity is the struggle against the background.

The double beta active nuclides are stable against normal beta decay but have the two-neutrino double beta decay ($2\nu 2\beta$) channel open: $Z\rightarrow(Z+2)+2e^- +2\bar{\nu}_e$. This last is a very rare, but standard, second order weak process and happens if the ground level of the Z isotope is lower than that of Z+1 but higher than that of Z+2. For massive majotrinos the process $Z\rightarrow(Z+2)+2e^-$, the $0\nu 2\beta$ decay, can take place. The relevant diagram is shown in Fig. 4a

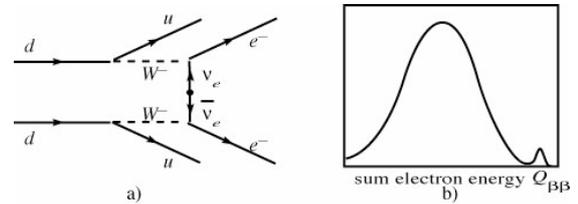

Fig. 4. a) $0\nu 2\beta$, b) Sum energy spectrum

There are two basic kinds of experiments[22]. The first is calorimetric; the source and the detector coincide and one measures the total energy released in the decay by the two electrons. Ideally, a spectrum as shown in Fig. 4b is expected: continuous for the $2\nu 2\beta$ decay, where some energy is taken by neutrinos, a single line (height exaggerated in the figure) at the transition energy ($Q_{\beta\beta}$) for $0\nu 2\beta$, where all the energy goes to the electrons (typically $Q_{\beta\beta}$ = 1-2 MeV). In practice the spectrum is superimposed on the background. To fully exploit the advantage given by the mono-chromaticity of the signal, detectors must attain very good energy resolution (a few keV), which must be coupled to extremely low background conditions.

In the second type of experiments the source is a sheet of the active metal, thin enough to allow the electrons to exit and be detected in the surrounding tracking chambers. The charges of the electrons and their momenta are measured in a magnetic field. The pros are the very clear signature and the possibility to use several different isotopes, the cons are the relative smallness of the source mass and, given the large size of the apparatus, the increased difficulty in sealing out backgrounds as those due to Rn. But the main problem is that, to get rid of the ultimate background, the tail of $0\nu 2\beta$, one needs, at, say, the 10 meV sensitivity, an energy resolution of a few keV. This should be compared with the present 90 keV of NEMO3[23], the best experiment of this type. For reasons of space I will limit the discussion to the calorimetric techniques and, amongst these, the two most sensitive: the Te bolometers of CUORICINO[24] and CUORE[24] and the enriched $^{75}$Ge diodes of Heidelberg-Moscow (HM)[21] and GERDA[25] at LNGS and MAJORANA[26] proposed in the US.

The observation of a $0\nu2\beta$ would prove the Majorana nature of neutrinos. To go further, we must extract the "effective mass"

$|M_{ee}|=||U_{e1}|^2 m_1 + |U_{e2}|^2 m_2 e^{i\alpha} + |U_{e3}|^2 e^{i\beta} m_3|||$

from the observed rate. For this the relevant nuclear matrix elements must be known, which are uncertain typically by factors ≥3. It is then mandatory to search on different double-beta active isotopes. Much more theoretical effort, joined to experiments aimed to measure critical quantities, is clearly also needed to reduce the theoretical uncertainties.

In presence of (Gaussian) background, the sensitivity of an experiment is given by, $F_M \propto [(t \times M/(\Delta E \times b)]^{1/4}$, where $b$ is the background index, $M$ the sensitive mass, $t$ the exposure time and $\Delta E$ the energy resolution. Unfortunately, it improves only with the 4$^{th}$ root of the exposure.

The most sensitive experiment is HM, now concluded, which ran at LNGS for 13 years, integrating an exposure of 71.7 kg yr. In 2002[21], part of the collaboration reported positive evidence of the signal with a claimed 4σ significance at the expected position $Q_{\beta\beta}$=2038.99±0.75 keV. The background index is $b$=0.2/(kg keV yr) before pulse shape analysis, 0.06 after, the energy resolution is 3.27 keV F.W.H.M. The signal is 28.8±6.9 events over a background of approximately 60 events, corresponding to a half-life $T_{1/2}$=(0.3–2)×10$^{25}$ yr, where the uncertainty is that of the matrix element. This corresponds to $|M_{ee}|$=(100-900) meV. Notice that the very good knowledge of $Q_{\beta\beta}$ and the superior energy resolution imply that the relevant backgrounds are only those in a narrow (say 60 keV) window around $Q_{\beta\beta}$. The background model obtained via Monte Carlo simulations contains a flat component and four lines of $^{214}$Bi. It fits the data reasonably well, but the positions of the Bi lines are off by a couple of standard deviations each.

The experiment with closest sensitivity, IGEX[27], again with enriched Ge, gives only the upper limit, $|M_{ee}|$=(100-900) meV ≤(330-1300) meV all the other experiments are even less sensitive.

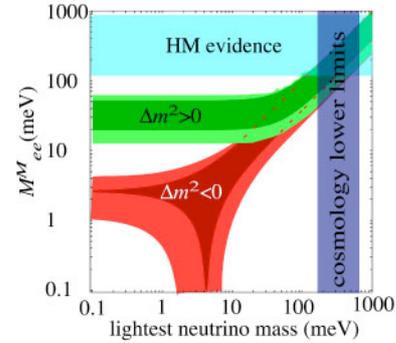

Fig. 5 Majorana effective mass.

Fig. 5 shows the expected value of $|M_{ee}|$ as a function of the lightest neutrino mass, for normal and inverse hierarchies. It is calculated[28] taking into account the oscillation data, for the nuclear matrix element of Klapdor et al.[29]. The darker bands correspond to the (complete) uncertainty on the Majorana phases, lighter colour bands include uncertainty on the other mixing parameters. The HM claim corresponds to a degenerate neutrino spectrum. The three, almost equal, neutrino masses are close to the cosmological limit (or even larger than the lowest ones). The result must be confirmed by future experiments. If this will not be the case a sensitivity down to a few 10 meV may be needed, a very difficult task indeed. The key features are background reduction and energy resolution. Working with "zero" background in the signal window and in the exposure time, the sensitivity goes with the root of the exposure, $F_M \propto [(t \times M/(\Delta E \times b)]^{1/4}$

The GERDA[25] proposal at LNGS is based on these concepts. The experience of Heidelberg-Moscow experiment has shown that Ge is one of the radio-cleanest materials and that residual backgrounds are largely located outside the detectors. It looks possible to aim at a background index $b$=10$^{-3}$/(kg keV yr) that would lead to a zero background exposure of a few 100 kg yr.

Heuser[30] in 1995 and Klapdor-Kleingrothaus et al. [31] in 1997 (GENIUS proposal at LNGS) have proposed to operate naked Ge crystals in liquid N2, taking advantage of the techniques developed by BOREXINO too produce extreme radiopurity (10$^{-16}$ g/g) liquid nitrogen. The GENIUS-TF[32] prototype at LNGS has shown that the concept is viable. See also the GEM proposal[33] in 2001 along similar concepts.

GERDA has further developed the idea designing a graded structure with a number of screening materials. The experiment foresees three phases. In the first the existing enriched Ge crystals of HM and IGEX, 17 kg, will be used at design background indices of $b=10^{-3}$/(kg keV yr) externally, and $b=10^{-2}$/(kg keV yr) internally. If the claimed signal is true this phase will confirm it in one year. The next phase aims to the sensitivity $M_{ee}$=(100 – 300) meV developing new detectors produced with minimisation of the cosmogenic activity. The third phase should improve by another order of magnitude, requiring a very strong effort. The idea is to join forces at this stage with the MAJORANA collaboration, which, in the meantime is developing complementary ideas for background suppression.

The bolometric techniques have been developed mainly by Fiorini and his group. Presently, the CUORICINO[23] experiment is taking data with 41 $TeO_2$ crystals, 0.76 kg each, corresponding to a total $^{130}$Te mass of 14 kg. With the present background level of $b$=0.2/(kg keV yr) and the 7 keV energy resolution the sensitivity is in the range of a few hundred meV. With luck it may confirm the HM claim, but, due to the uncertainty in the matrix elements, it cannot disprove it.

The next step will be CUORE[24], made of 19 CUORICINO-like columns and a sensitive Te mass of 741 kg. Simulations show that $b=10^{-2}/(\mathrm{kg\cdot keV\cdot yr})$ can be achieved. In this case in five years running time $M_{ee}$=(30 – 70) meV can be reached.

## 5. Conclusions and outlook

Neutrino physics made enormous progress in the last several years, showing for the first time physics beyond the Standard Model. This is mainly due to experiments using natural neutrino sources in underground laboratories. Perspectives are exciting, but the next round of experiments will be expensive and engaging, requiring resources that are large when compared with past generation $0\nu2\beta$ experiments, but still very limited in comparison with a typical collider experiment. But, the stake to find the origin of neutrino mass does not appear to me to be much less important than discovering the Higgs.

We must design a global road map, with complementarities amongst the regional programs and amongst accelerator and underground experiments and distinguishing between experiments that are feasible now and enterprises that are extremely expensive and useful only if unknown physics turns out to be kind enough ($\theta_{13}$ not too small)